\documentclass{elsart41}
\usepackage{graphics}
\usepackage{graphicx}
\usepackage{amssymb}

\begin{document}

\begin{frontmatter}



\title{Fulde-Ferrel-Larkin-Ovchinnikov State due to Antisymmetric 
Spin-Orbit-Coupling in Noncentrosymmetric Superconductivity  CePt$_3$Si}

%
\author[a]{Hiroko Tanaka},
\author{Hirono Kaneyasu},\,
\author{Yasumasa Hasegawa}

\address{Department of Material Science, University of Hyogo, 3-2-1 Kouto,
Kamigori-cho, Hyogo, Akou-gun 678-1297, Japan}

\corauth[a]{Corresponding author. E-mail : ri05i022@stkt.u-hyogo.ac.jp}

\begin{abstract}
When the inversion symmetry is broken, the spin-orbit coupling reduces the transition temperature
of some types of spin triplet superconductivity,
which is similar to the case that magnetic field reduces the spin singlet superconductivity 
due to Zeeman splitting.  
It is well known that  Fulde-Ferrel-Larkin-Ovchinnikov (FFLO) state of spin singlet superconductivity
 is realized 
near the Pauli limit (or Chandrasekhar-Clogston limit) of external magnetic field.
In FFLO state the amplitude of the order parameter is not uniform in space. 
In this paper we study the FFLO state in the spin triplet superconductivity in the absence of magnetic field
due to the spin-orbit coupling. Although the FFLO state is not realized in the simple model 
with spherical Fermi surface, it will be stabilized if some condition is favorable for it.
We discuss the possibility of FFLO state in CePt$_3$Si in the absence of external magnetic field.

\end{abstract}

\end{frontmatter}
In CePt$_3$Si the upper critical field $H_{c2}$ exceeds the paramagnetic
limiting field\cite{Bauer}, which is thought to be an evidence of the spin triplet pairing.
Recent experiments on Knight shift\cite{Ueda,Yogi2006} also indicate the spin triplet superconductivity.
%
%
This seems to conflict with a general statement
that spin triplet pairing is forbidden in the system with broken 
inversion symmetry\cite{Anderson}.  
Frigeri et al.\cite{Frigeri} 
studied the effect of the Rashba type spin-orbit coupling, 
\begin{equation}
  H_p = \alpha \sum_{\mathbf{k}, s,s'} \mathbf{g}_{\mathbf{k}} \cdot \mathbf{\sigma}_{s,s'}
   c_{k,s}^{\dagger} c_{k s'} ,
\end{equation}
with $\mathbf{g}_{\mathbf{k}}=(-k_y, k_x,0)$
as a model for CePt$_3$Si
and obtained that the spin triplet pairing is not destroyed by the spin-orbit coupling, 
if the $\mathbf{d}$-vector satisfies $\mathbf{d}(\mathbf{k}) \parallel \mathbf{g}_{\mathbf{k}}$.
This state is not consistent with the line nodes of the energy gap observed by the experiments of
penetration depth\cite{bonalde} and thermal conductivity\cite{izawa}. 
In order to explain the existence of the line nodes, the mixing  state of triplet and singlet 
pairings has been studied\cite{hayashi}.

In this paper we study the  FFLO state caused the pair breaking due to
 the spin-orbit coupling in the system without inversion symmetry.
We calculate $T_c$ as a function of $\alpha $
for the 
Cooper pairs of  $(\mathbf{k} +\frac{\mathbf{q}}{2}, s_1)$ 
and $(-\mathbf{k}+\frac{\mathbf{q}}{2},s_2)$.


%
%

The linearized gap equation for FFLO state is written in a $2 \times 2$ matrix form\cite{Frigeri} as
\begin{eqnarray}
\Delta_{ss'}&(&\mathbf{k} +\frac{\mathbf{q} }{2}) 
=-k_{B}T_{c}\sum_{\mathbf{k}' ,n}\sum_{s_1,s_2}V_{\mathbf{kk} '}
G_{ss_1}^{0}(\mathbf{k}' +\frac{\mathbf{q}}{2},i\omega _{n}) \nonumber \\
& &\times\Delta_{s_1s_2}(\mathbf{k}' +\frac{\mathbf{q}}{2})G_{s's_2}^{0}(-\mathbf{k}' 
+\frac{\mathbf{q}}{2}',-i\omega _{n}),
\end{eqnarray}
where 
\begin{equation}
G^{0}(\mathbf{k}, i\omega _{n})=
G_{+}(\mathbf{k}, i\omega _{n})\sigma _{0}+
(\hat{\mathbf{g} }_{\mathbf{k}}\cdot \mathbf{\sigma} )
G_{-}(\mathbf{k}, i\omega _{n}) ,
\end{equation}
\begin{equation}
G_{\pm }(\mathbf{k}, i\omega _{n})
=\frac{1}{2}[(i\omega _{n}-\epsilon _{\mathbf{k},+})^{-1}
\pm (i\omega _{n}-\epsilon _{\mathbf{k},-})^{-1}] ,
\end{equation}
$
\epsilon _{\mathbf{k}+\frac{\mathbf{q}}{2},\pm }
=\xi _{\mathbf{k}+\frac{\mathbf{q}}{2}}\pm \alpha \mid \mathbf{g}_{\mathbf{k} +\frac{\mathbf{q}}{2}}\mid ,
$
and
$\sigma _{0}$ is the unit matrix.

The gap function is decomposed into a spin singlet part 
and a spin triplet 
part,   
$\Delta (\mathbf{k} +\frac{\mathbf{q}}{2})
=[\psi (\mathbf{k}+\frac{\mathbf{q}}{2})\sigma_{0}+\mathbf{d}(\mathbf{k}+\frac{\mathbf{q}}{2})
\cdot {\mathbf{\sigma }}]i{\sigma }_y$.
%
%
Spin triplet part and spin singlet part are mixed in general.
When the particle-hole symmetry is satisfied, however, the the singlet and triplet order parameters do not mix.
%
We set $\mathbf{q}=(0,0,k_{F}\rm{q})$ and $\xi _{\pm \mathbf{k}+\frac{\mathbf{q}}{2}
}=\xi _{\mathbf{k}}\pm \frac{\hbar ^{2}k_{F}^{2}\rm{q}\rm{cos}\theta }{2m}$.
Then we find the transition temperature for spin triplet pairing is given by
\begin{eqnarray}
\ln  &(&\frac{T_c}{T_{c0}}) 
=\pi k_{B}T_{c} 
\langle {\rm Im}\sum_{n=0}^{n_{c}-1}
\big( \nonumber \\
&(&f_{1-}+f_{2-}+f_{2+}+f_{1+} -\frac{4}{2i \left| \omega _{n} \right|}) 
{\left| {\mathbf{d}(\mathbf{k}+\frac{\mathbf{q}}{2}) }\right| }^2 \nonumber \\
&-&(f_{1-}-f_{2-}-f_{2+}+f_{1+}) \nonumber \\ 
&\times&[2(
\hat{\mathbf{g}}_{-\mathbf{k}+\frac{\mathbf{q}}{2}}
\cdot \mathbf{d} (\mathbf{k}+\frac{\mathbf{q}}{2}))
(\hat{\mathbf{g}}_{\mathbf{k}+\frac{\mathbf{q}}{2}}
\cdot \mathbf{d}(\mathbf{k}+\frac{\mathbf{q}}{2})) \nonumber \\ 
& &-\hat{\mathbf{g}}_{-\mathbf{k}+\frac{\mathbf{q}}{2}}
\cdot \hat{\mathbf{g}}_{\mathbf{k}+\frac{\mathbf{q}}{2}}
\left| \mathbf{d}(\mathbf{k}+\frac{\mathbf{q}}{2})\right| ^2 ]
\big)
\rangle_{\mathbf{k}}.
\end{eqnarray}
where 
\begin{eqnarray}
f_{1\pm }=\big(\frac{\hbar ^{2}k_{F}^{2}\rm{qcos}\theta }{m}&+&2i\mid \omega _{n}\mid \nonumber \\
&\pm &\alpha (\mid {\mathbf{g}_{\mathbf{k}+\frac{\mathbf{q}}{2}}}\mid 
-\mid {\mathbf{g}_{-\mathbf{k}+\frac{\mathbf{q}}{2}}}\mid )\big)^{-1}, \\
f_{2\pm }=\big(\frac{\hbar ^{2}k_{F}^{2}\rm{qcos}\theta }{m}&+&2i\mid \omega _{n}\mid \nonumber \\
&\pm &\alpha (\mid {\mathbf{g}_{\mathbf{k}+\frac{\mathbf{q}}{2}}}\mid 
+\mid {\mathbf{g}_{-\mathbf{k}+\frac{\mathbf{q}}{2}}}\mid )\big)^{-1},
\end{eqnarray}
and $T_{c0}$ is the transition temperature for $\alpha=0$.


We calculate $T_c/T_{c0}$ as a function of $\alpha$ for several choice of order parameters;
$\mathbf{d}(\mathbf{k})=\hat{\mathbf{x}}k_{x}+\hat{\mathbf{y}}k_{y}+\hat{\mathbf{z}}k_{z}$,
$\mathbf{d}(\mathbf{k})=\hat{\mathbf{x}}k_{x}+\hat{\mathbf{y}}k_{y}$ 
and
$\mathbf{d}(\mathbf{k})=\hat{\mathbf{x}}k_{x}^3+\hat{\mathbf{y}}k_{y}^3$ 
The transition temperature is suppressed by $\alpha$ in these states. 

\begin{figure}[tbh]
\begin{center}
\includegraphics[width=0.5\textwidth]{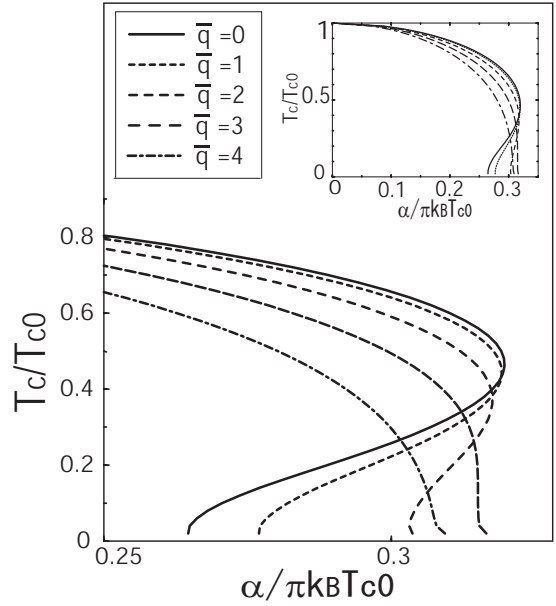}
\end{center}
\caption{ Transition temperature  as a function of $\alpha $ 
for triplet superconductivity with $\mathbf{d}(\mathbf{k})=\hat{\mathbf{x}}k_{x}+\hat{\mathbf{y}}k_{y}$
and
$\mathbf{g}_{\mathbf{k}}=(-k_{y}, k_{x}, 0)$.}
\label{fig1}
\end{figure}
In the Fig.~1, we plot transition temperature of the state $\mathbf{d}(\mathbf{k})=(k_x,k_y,0)$ 
as a function of $\alpha$ for some fixed values of $\bar{\rm q}$, 
where we use $\mathbf{g}_{\mathbf{k}+\frac{\mathbf{q}}{2}}=(-k_{y},k_{x},0)$,
$\mathbf{d}(\mathbf{k})=\hat{x}k_{x}+\hat{y}k_{y}$,
and define $\bar{\rm q}$ by
$\bar{\rm q}=\frac{(\hbar ^{2}k_{F}^{2}/m\pi k_{B}T_{c0})}{\alpha /\pi k_{B}T_{c0}}\rm q$.
Since we set $\mathbf{q} \parallel \hat{z}$,
$\mathbf{g}_{\mathbf{k}+\frac{\mathbf{q}}{2}}$   
is independent of $\mathbf{q}$. 
For $\bar{\rm q}=0$, the transition temperature shows a reentrance near 
$\alpha/(\pi k_B T_{c0}) \approx 0.3$
(Fig.~1). This curve is similar to the transition temperature of the singlet
 superconductivity as a function of the magnetic field,
where the only the Pauli pair breaking effect of the magnetic field is considered.
The reentrance never occurs, since the transition curve is calculated by assuming the second order transition. 
When the temperature is lowered at fixed $\alpha /\pi k_{B}T_{c0} \approx 0.3$,
the order parameter is finite below the higher transition temperature 
and the 
lower transition temperature is spurious.
In the case of Pauli pair breaking, the FFLO state becomes stable 
for the magnetic field higher than the critical value.
We had expected the similar situation for the case of pair breaking due to the spin-orbit coupling.
As seen in Fig.~1, however, we could not find the region where FFLO state 
becomes a ground state through a second order transition.
In this study we assume the spherical Fermi surface for simplicity. If we consider the Fermi surface 
with different shape with lower dimensionality or better nesting condition, FFLO state is expected to appear
in some parameter region, as is the case 
in the ordinary FFLO state caused by the Pauli pair breaking\cite{Shimahara}.

In conclusion,
we study the FFLO state of the triplet superconductivity due to the spin-orbit coupling. 
Although we did not find the FFLO phase similar to an conventional FFLO one,
we expect that FFLO state will be stabilized for the system with suitable Fermi surface.
If the FFLO state is realized without magnetic field, the amplitude of the energy gap
varies in space and  the specific heat and penetration depth will
depend on temperature as  the power low.



We are very grateful to H. Shimahara for useful discussions.


%

\end{document}